\documentclass{emulateapj}

\slugcomment{To be published in PASP (January 2006)}

\shorttitle{IFS of Planet Transits}
\shortauthors{Arribas et al.}

\begin{document}


\title{Exploring the Potential of Integral Field Spectroscopy Observing Extrasolar Planet Transits: \\
    Ground Based Observations of the Atmospheric N\lowercase{a} in  HD209458b}


\author{Santiago  Arribas\altaffilmark{1-4}, Ronald~L.\  Gilliland\altaffilmark{1}, William~B.\ Sparks\altaffilmark{1}, Luis 
L\'opez-Mart\'in\altaffilmark{3}, Evencio Mediavilla\altaffilmark{3}, Pedro G\'omez-Alvarez\altaffilmark{3}}

\altaffiltext{1}{Space Telescope Science Institute, 3700 San Martin Dr., Baltimore, MD 21218, USA}
\altaffiltext{2}{Affiliated with the Space Telescope Division of the European Space Agency,  ESTEC, Nordwijk, Netherlands}
\altaffiltext{3}{Instituto de Astrof\'\i sica de Canarias, 38200 La Laguna, Tenerife, Spain}
\altaffiltext{4}{Consejo Superior de Investigaciones Cient\'\i ficas (CSIC), Spain}


\begin{abstract}

We explore the use of Integral Field Spectroscopy (IFS) for observing extrasolar planet transits. Although this technique should find its full potential in space observations (e.g.\ JWST, TPF), we have tested its basics with ground based time series observations of HD209458b obtained  with WHT + INTEGRAL during a transit in August 17/18, 2004.  For this analysis we have used 5550 spectra (from a potential of $\sim$30000), obtained in 150 exposures during a period of more than 7~hours.  We have found that IFS offers three fundamental advantages with respect to previously used methods (based on imaging or standard slit spectroscopy). First, it improves the effective S/N in photon limited observations by distributing the light coming from the star into the two dimensions of the detector. Second, this type of IFS data allows to {\it autocalibrate\/} instrumental and background effects. Third,  since the star image characteristics (i.e.\ seeing, spatial shifts, etc.) as well as its photometric properties are extracted from the same data-cube, it is possible to decorrelate photometric instabilities induced by PSF variations. 

These data have also allowed us to explore the accuracy limits of ground based {\it relative\/} spectrophotometry. This was done using a photometric index that probes the NaD lines, for which we obtained a nominal 1-$\sigma$ error of  $\sim$$1.0\cdot10^{-4}$. This result, based on observations of only one transit, indicates that this type of ground observation can constrain the characterization of the transmission spectrum of extrasolar planets,  especially if they cover multiple transits under good weather conditions.  The present observations are compatible with no extra NaD depression during the transit. Although this result seems to be inconsistent with the recently reported  HST -- STIS findings we point out its limited statistical meaning:   the results disagree within 1-$\sigma$, but agree within 2-$\sigma$. 

This method requires careful and specific reductions, and details about this process are given. We also give some recommendations to instrument developers in order to enhance the efficiency of the method.          

\end{abstract}





\section{Introduction}

The transit method (TM) has recently gained much attention among the various methods for the detection of extrasolar planets due to its potential for discovering earth-like planets. This method also offers the possibility for studying the existence of planetary satellites, rings, and atmospheric features (see, for instance, Deeg 2002). All these possibilities have been clearly demonstrated with HD209458, the first planetary transit discovered (Charbonneau et~al.\ 2000; Henry et~al.\ 2000; Mazeh et~al.\ 2000). The light curve of this object has allowed an unprecedented characterization of the stellar/planet system (e.g.\ planetary and stellar radii, stellar limb darkening, orbital inclination, period, etc), imposing limits to the presence of planetary satellites and rings (Brown et~al.\ 2001), and detecting features associated with the planet atmosphere (Charbonneau et~al.\ 2002; Vidal-Madjar et~al.\ 2003, 2004). 

The success of the TM largely relies on the (spectro) photometric accuracy for which the light curve can be obtained. For some science applications (e.g.\ detection of planetary atmospheric features) only accurate {\it relative} spectrophotometry (i.e.\ band1/(nearby)band2 and  in/out transit) is required, reducing drastically systematic correlated errors which may otherwise limit the S/N. Experiments with the HST (Gilliland, Goudfrooij, Kimble 1999; Brown et~al.\ 2001; Gilliland and Arribas 2003) have demonstrated that accuracies of $\sim$10$^{-5}$  or higher can be reached from space. 

Such a high accuracy is not always required. For instance in the wavelength domain 1190--1710~\AA, Lyman $\alpha$, OI, and CII lines suffer during the transit a drop of $\sim$15 , 13 and 7.5 percent, respectively (Vidal-Madjar et~al.\ 2003, 2004). In optical wavelengths, the variations are indeed smaller. Charbonneau et~al.\ (2002) have found a drop of 2.5 parts in 10000 (i.e.\ 0.025 percent) during the transit in the NaD resonant lines. This drop is smaller than expected from (poorly constrained) models, which  also predict that other spectral characteristics in the optical (e.g.\ Rayleigh continuum) should be detected with lower accuracies (e.g.\  Brown 2001).  The smaller than expected  NaD absorption detected by Charbonneau et~al.\ (2002) may have resulted from a high cloud deck, a low atomic Na abundance, or a combination of both. 

In this context,   it is relevant to understand which are the accuracy limits that can be achieved from the ground. This should allow us to know if part of future studies can be performed taking advantage of the more abundant and flexible observing time available from ground-based observatories, or if we need to rely entirely on space data. Note that, with the advent of large aperture ground telescopes, the Poisson limit may be boosted allowing access to more distant systems. This is particularly relevant in connection with follow up observations of the planetary transits to be discovered  by missions such as Kepler (Borucki et~al.\ 2004).  Pioneering attempts to detect spectral features due to the planetary atmosphere from the ground in 51~Peg and HD209458 include those of Coustenis et~al.\ (1998), Rauer et~al.\ (2000), Bundy \& Marcy (2000), Moutou et~al.\ (2001), and Brown et~al.\ (2002).  More recently Deming et~al.\ (2005) using NIRSPEC on the Keck~II telescope have  reported upper limits to the CO absorption at a sensitivity level which suggests that a general masking mechanism is present in the planetary atmosphere of HD209458b. Also recently Narita et~al.\  (2005)  have studied the transmission spectrum of this object on the basis of high resolution spectroscopy with {\it Subaru\/}. Although the accuracy of these observations were unable to confirm or contradict the result of Charbonneau et al., they reported upper limits on absorption due to several optical transitions. All these  ground based observations have gained importance since STIS--HST  is no longer operational. 

In this paper we propose to  take advantage of the comprehensive collecting nature of IFS to perform this type of observation. The observational method, whose advantages are described in Section~2,  is tested with ground based data. The observations are described in Section~3. Details about the reduction process for this particular application are described in Section~4. The results are shown and discussed in Section~5, while Section~6 summarizes the main conclusions.  

\section{ Proposed Method: The potential of IFS in observations of planet transits}

Gilliland, Goudfrooij, and Kimble (1999) proposed to increase the photometric accuracy of very high S/N observations by using the spectroscopic mode of STIS. By distributing the star light along the dispersion direction it is possible to increase the total number of photons collected before the detector reaches the saturation limit. This simple idea enhances the duty cycle of the instrument  (i.e.\  total number of photons by unit of available time, including overheads), which translates to an increase in the S/N when the photon noise is the major source of noise. The larger detector footprint of the stellar light has a negligible consequence in terms of adding noise in a very high S/N regime. Actually it has the beneficial effect of averaging out flat field and sensitivity residuals.  Another practical advantage is the possibility to define {\it a~posteriori\/} ad hoc 'filters' well suited for the science applications.  

Here we propose to extend this idea by means of integral field spectroscopy  (IFS). Thanks to the 2D $\rightarrow$ 1D conversion done by a fiber bundle (or by an image slicer),  the star light is not only spread along the dispersion direction but also across dispersion. This technique allows us to use a larger area of the detector, boosting the limit imposed by the photon-noise per exposure. Note that to optimize the use of the detector, the image of the star at the focal plane should be well oversampled by the Integral Field Unit (IFU). Since for these transit observations we are interested in the total number of photons recorded in a limited period of time (i.e.\ duration of the transit), the image at the input of IFU could be defocused. Note that defocussing the star image on the IFU does not modify the spectral resolution of the observations, which is another practical advantage of the IFS method with respect to previously used spectroscopic methods.   

The ability to distribute the light over the whole detector is particularly relevant with the advent (and prospects) of very  large telescopes. This can be easily understood with the following example. Typical imaging observations of HD209458b (V~= 7.64) in a 100~\AA\ band with an 8m.\ telescope will saturate in $\sim$0.1~sec. Considering a typical readout time of 60~sec, and taking into account that the transit lasts $\sim$3~hours, an effective integration time of $\sim$20 seconds on the source during the transit could be obtained. If the observations are now done with a standard slit spectrograph working at, say, 1~\AA/pixel,  saturation is roughly reached in $\sim$10~sec, and the effective integration time becomes  $\sim$30~minutes. If now the observations are done with an IFS at the same spectral resolution and sampling the (defocused) star image with, say, 100 spatial elements on the IFU ({\it spaxels\/}, according to IFS community), the effective exposure time may be increased up to 2.8~hours (out of 3 available). Of course, real observations can be optimized by defining faster readouts, modifying the spectral resolution, etc. Detectors like that used by ULTRACAM (Dhillon \& Marsch 2002) may also reduce overheads. The numbers above are just  illustrative of current standard detectors. 

Apart from the ability to gather photons, stability is crucial for TM observations. Of course, beyond some limits it is not possible to control the stability of the environment and/or the observational set-up. While this is true for both space and ground observations, the unstable nature of the Earth atmosphere makes very accurate ground observations particularly challenging. In any case, when uncontrolled instabilities may affect the photometric accuracy of the system, a good record of them may give the possibility to decorrelate the photometric signal. This provides another important advantage of the IFS. Since both, the spectra from which the photometric information is obtained {\it and\/} the images of the object (PSFs) are extracted from the same data cube, we can remove photometric variations induced by PSFs instabilities. Furthermore, in fiber-based IFS (and to a lesser degree a slice-based IFU) a shift of the image on the focal plane may be well tracked independently of a shift between the pseudo-slit and the detector. This gives full control of the PSF variations, as well as the optical and mechanical shifts during time series observations.  Fiber systems also provide a series of generic advantages such as, for instance, the azimuthal scrambling of the light (which reduces errors in the barycenter of the recorded fiber image), the possibility to use static spectrographs (which reduces flexure induced errors), etc (see more details in, for instance, Arribas \& Mediavilla, 2000).  On the other hand, potential problematic aspects of fiber-based IFS systems are the modal noise due to small changes in fiber position or illumination (e.g.\ Baudrand and Walker 2001), and a possible temporal variation of the fiber throughput with wavelength. However, we will see later that, at least for the present instrumental configuration, our data do not suggest they are significant sources of noise.

Another advantage of IFS for these type of observations is a consequence of the large number of spectra collected simultaneously, which can be used to {\it autocalibrate\/} the data themselves from detector and background signatures. This will be shown in Section 5. 

Summarizing, IFS provides three main advantages for transiting planets observations: i)~improves the ability to collect photons during the transit, enhancing  the S/N, ii)~it allows to autocalibrate the data, and iii)~by tracking in an independent manner the instabilities produced at the focal plane and those due to the spectrograph, it is possible to remove noise correlated to PSF characteristics.

\section{Observations}


We performed the observations with the 4.2m William Herschel Telescope during 2004 August 17/18 using the INTEGRAL system (Arribas et~al.\ 1998) connected to WYFFOS (Wide Field Fiber Optic Spectrograph; Bingham et~al.\ 1994).  We used the recently installed WYFFOS long camera, equipped with two EEV-42-80 thinned and coated CCDs butted along their long axis to provide a 4K $\times$ 4K pixel mosaic.  The high speed read out mode  (45~sec) was used, with gains of 2.3 e- (chip1-red) and 1.94 e- (chip2-blue) yielding a readout noise of 6 e-.  These detectors are considered linear for standard observations until $\sim$52~K ADU (see below). Note that WYFFOS is mounted in  one of the Nasmyth platforms of the WHT, which provides good mechanical stability (see below).

The bundle of fibers SB1 (Standard Bundle~1), consisting of 205 fibers, each 0.45$''$ in diameter, was used. This bundle is arranged such that 175 fibers cover a rectangular area of 7.8$'' \times6.4''$  while 30 additional fibers, forming a ring of 90$''$ in diameter concentric with the rectangle, measure simultaneously the sky background.  The actual distribution of the fibers at the focal plane can be found in Figure~6 of Arribas et~al.\ (1998) and also in Figure~1 of del Burgo et~al.\ (2000).  

The spectra  were taken using the R316R line/mm grating with an effective resolution of $\sim$3~\AA\ and covering the 
$\sim$4000--10000~\AA\ spectral range. 

INTEGRAL does not have an independent focus mechanism for the fiber bundles, so the only possibility to spread the light over the input of the fiber bundle in a controlled way was to defocus the telescope. However we decided not to apply any major defocus to the telescope, because it would have affected the guiding quality. 

The observations on target consisted of 8 time series of 20 individual exposures of 120 sec each (the 8th series had to be suspended after 10 individual exposures were taken due to the detectable flux increase due to the twilight). After each time series we took a short flatfield exposure. Therefore, the total object data set consisted of 150 individual exposures (with a potential of 
$\sim$30000 individual spectra though, of course, many had low S/N), extending for a period of more than 7~hours, from which 
5~hours were integrating on the target.  The weather conditions were good, with a stable seeing of $\sim$0.8~arcsec. The values of atmospheric extinction in the visible ($r'$-band) obtained by the Carlsberg Meridian Telescope (http://www.ast.cam.ac.uk/$\sim$dwe/SRF/camc$\_$extinction.html) also suggest stable and photometric conditions for the night, with a mean value for the extinction coefficient  of $0.091\pm0.011$ magnitudes per airmass using data expanding a period  of 4~hours. Non-photometric data were not detected this night.

We located the object slightly off-center (i.e.\ around fiber number 111 in the standard code for this bundle) in order to minimize the effects of one broken fiber located in the opposite direction. We will refer to it as {\it central fiber\/} or {\it fiber~111}, and it is the closest to the peak of the image in most of the exposures. The 37~fibers around it covered most of the target flux.  

Apart from lamp-flats taken between sets of 20~exposures, bias, wavelength calibration, and sky flat images for calibration purposes were taken at the beginning and at the end during the twilight.  The internal calibration illumination does not mimic the exit pupil of the telescope, producing slightly larger fiber images. This should be irrelevant provided the use of the these calibration images for the present application (see next section).




\section{Reductions}

The reductions of these IFS data have the goal of generating both spectra (from which the relative spectrophotometric index is derived) and images (from which PSF characteristics are obtained) for each of the 150 individual exposures. Here we will briefly describe the steps followed during this process, commenting on the relative importance of each step for this particular application. 

\subsection{Spectral reductions}

The present study will be focussed on the analysis of the NaD lines. As mentioned in Section~1 these lines have been studied with great accuracy with HST (Charbonneau et~al.\ 2002), which provide us with a  good reference to compare the present observations. For this reason, and with the aim of simplifying and making the reduction process more manageable, we trimmed our original images into a spectral region of $\sim$200~\AA\ including the above mentioned resonant lines.

\subsubsection{Bias:}  

The bias value was obtained from the overscan strips signal. For the red chip used here, the mean value in that region increased linearly over the night from 1018 to 1028~ADUs. The rms of this relation is about  $\la$ 1~ADUs, a very small value compared with the typical signal of the brightest spectrum ($\sim$40000~ADU). This indicates that the expected errors associated with an incorrect bias subtraction should have a minimal effect on the relative photometry of these very high signal to noise observations.

\subsubsection{Flat Field correction:} 

By construction, in many IFS systems is not possible to illuminate directly the detector with a uniform source (e.g.\ blank sky). This is due to the presence of the IFU (i.e.\ fiber bundle, image slicer, microlens array) in the optical path.  This prevents obtaining directly flats fields for correcting pixel to pixel sensitivity variations. Instead  {\it flats\/}  usually refers to images obtained when illuminating uniformly the input of the IFU, and sometimes are called {\it flat-spectra\/}.  These are generally used to correct both, differences in throughput among the different  sampling elements (i.e.\ fibers), and sensitivity variations of {\it extracted pixels\/}. Note that an extracted pixel is generally obtained by integrating several individual pixels across the dispersion direction, so they represent an averaged (weighted) behavior. Although this approach is generally enough for most applications, in some cases it may represent a limitation.  We will show below that our spectra drift about 0.6~detector pixels during the 7.5 ~hours of the observations, so different pixel sensitivities could emulate object flux variations. 

Although it is not possible to obtain direct detector-flats  with INTEGRAL/WYFFOS, its detector is also used at the prime focus of the WHT for direct imaging. Fortunately this detector was  used for several imaging programs during September 2004. So we collected the flats images recorded during this period (kindly provided to us by D.~Lennon) to create a very high S/N {\it superflat\/}. For the spectral range considered here we combined about 40 images (mostly in the V filter) and corrected them  from low spatial frequency variations using the IRAF tasks {\it imcombine\/} and {\it flat1d\/}, respectively.  Note that although relatively large fringing amplitudes have been measured at longer wavelengths ($\lambda>7000$~\AA), the wavelength range analyzed in this paper is free of that effect. 
 This superflat has about  $10^6$~ADU ( $2.3\cdot10^6 ~e^-$) and, therefore, a nominal S/N per pixel of $\sim$1500.

\subsubsection{Linearity correction: }

The photometric index at a given time is obtained after averaging the individual values obtained from the spectra recorded at that particular time. The fact that the relative intensity of these individual spectra may change during the night (due not only to guiding and seeing instabilities, but also to the unavoidable differential atmospheric diffraction), makes the spectral index sensitive to detector linearity deviations. 

The linearity of the present detectors was studied by Tulloch (2001). We have fit a polynomial function to his experimental data with signals $\la$\,38~K~ADU (rms of 0.14 percent). For signals $>$38~K~ADU, the slope of the linearity curve changes sharply from its relatively smooth behavior at lower signals. This slope change is difficult to model in detail,  so we decided not to correct for this linearity slope change (i.e. we apply a constant correction for values $>$38~K~ADU).  Linearity corrections at the lower signal end (i.e.\ $<$1000~ADU) are also difficult to model with high accuracy, but effects on the relative photometry are small.  In Section~5, we will analyze and correct for the residuals of this linearity correction.

\subsubsection{Extraction:}

The detector readout noise has a minimal effect in these very high signal to noise observations, we therefore decided to use relatively wide extraction windows of 14~pixels. This width guarantees the collection of most of the flux from a fiber/spectrum, but still avoids overlapping information from different spectra.  In addition, taking into account the relatively small drift of the spectra in cross dispersion over the night ($\sim$0.6~pixels in 7.5~hours), we decided to use a fixed extraction window for all the spectra/exposures. (However, we also reduced the data allowing for global shifts in the location of the extraction windows, but no appreciable changes were detected.) The functions defining the centers of the extraction windows were obtained by adjusting polynomials to the spectra of the image obtained by adding all the flat-lamp exposures (after previous removal of the cosmic rays). It was checked that these spectra were actually located near the mean position during the night.  These functions were 5th degree polynomials obtained after fitting 130 data points along the selected spectral range.  Typical  rms were of 0.015 pixels (no significant improvement was found when using higher degree polynomials). The extraction was done as a simple summation (i.e.\ not with PSF shape and inverse variance weighting) in order to avoid undesired renormalization factors which could affect the flux.  Although these factors should have a very small effect in standard observations, one could think that they may have some relevance for the current ultra-precise photometry. Therefore, we decide not to use this approach which, in addition, has very little advantage in a high S/N regime. In any case, we checked that no relevant differences were found when an optimal extraction was performed. 

\subsubsection{Wavelength calibration:}

 For the present analysis the wavelength calibration is important to define accurately the bands used for creating the spectral index.  Although the absolute conversion from pixels to wavelength coordinates does not need to be particularly accurate, it is important to correct any drift or change of the wavelength solution among fibers and, especially, during the night. 

For obtaining the absolute wavelength calibration for each fiber we used the wavelength calibration exposures obtained for such purpose.  That was done using as reference the brightest spectrum with the IRAF task {\it calibrate\/}. Then, the individual wavelength solutions for each fiber were obtained with {\it recalibrate\/}.   To detect any drift along the dispersion direction we cross-correlated the three brightest spectra in the images obtained along the night. These spectra are located in the region of the detector where most of the signal is extracted.  We find a similar behavior in the three cases with a linear shift of $\sim$0.3~pixels over the night at a constant rate of  $\sim$0.04~pix/h, with a typical rms of 0.05~pixels. Therefore, to improve the accuracy of the drift  associated with each exposure we combined all the spectra obtained in a given exposure, and cross-correlated these integrated spectra.  

Once the wavelength solution (and spectral drift) was known, we decided not to resample the individual spectra to a common spectral range and dispersion. This step would have implied an extra interpolation of the data. We found it preferable to use this information to define homogeneously the spectral bands of all the spectra (see Section~5.2). The wavelength solution was applied with the task 
{\it dispcor\/} of IRAF.

\subsubsection{Cosmic ray rejection:}

The cosmic ray rejection is an important step in the reductions, since a relatively small number of badly removed cosmic rays could degrade substantially the S/N of the data. After trying several automatic rejection methods we decided that none of these methods was sufficiently reliable for our particular case. The reason for this is that, because the spectra are well separated on the detector, slight shifts of the image at the focal plane may produce relatively important changes in the flux at the detector, which may be interpreted as CR when different exposures are compared.  Therefore we decided to follow the secure method of removing the CR in an interactive manner. In order to identify more easily cosmic rays, each individual spectrum was divided by a high S/N clean template, so even if a CR hits in the middle of the spectral feature (i.e.\ the NaD lines)  it could be removed reliably. After the cosmic rays were identified and removed the spectra were multiplied by the template.  A total of $\sim$80 cosmic rays were identified and removed in the relevant spectral range but, of course, the full set of over 5000 spectra were inspected. 

\subsubsection{Throughput correction:}

Flat-spectra were applied to correct for fiber sensitivity. This was done using the blank sky observations during the twilight at the beginning and at the end of the night.  Although the fiber-to-fiber sensitivity variations are large (rms $\sim$20~percent) these are stable during the night, with typical differences of smaller than $\sim$1~percent (i.e.\ the difference of two flats taken at the beginning and at the end of the night gave a rms of 1.1~percent for the 37~fibers used, which probably imposes an upper limit to the actual fiber throughput stability).   Note that global fiber sensitivity variations have no impact on the determination of spectral indices, which always probe relative flux variations (i.e.\ band1/band2).  A temporal variation in the fiber throughput with wavelength could be a limitation, but we don't have evidence for such a behavior. (As we will see in Section~5, changes in the throughput-wavelength function seem to be dominated by atmospheric more than for instrumental effects).  

Although the photometric index is not dependent on the correction of fiber throughputs, this affects  the star image reconstruction. In fact errors in the fiber throughput corrections may produce incorrect  PSF  reconstruction, introducing further uncertainties during the decorrelation process.  The throughput correction may also affect the background subtraction. However, in these two cases the effects on the final results are very small (see below). 

\subsubsection{Background subtraction:}

As described in Section~3,  30~fibers located in a ring of 90$''$ in diameter around the star allowed us to obtain contemporaneous background information. The spectra of some of these fibers are located relatively close to fibers with very high signal,  so they may be affected by cross-talk (or scattered light). Therefore, the averaged value of only 18 of these fibers was actually used to obtain the background information.  

\subsection { Reconstructed images}

In order to obtain the star images, we created files with the X and Y~positions of the fibers in the focal plane of the telescope and the integrated  (background-subtracted) flux in the range 5829.200--5957.321~\AA\ used for the definition of the photometric index  in Section~5 (i.e.\ $f_t = f_1 + f_2 + f_3$). With the help of the NAG routine E01SAF, an interpolating 2-dimensional surface F(x, y) is generated. This routine guarantees that the constructed surface is continuous and has continuous first derivatives. The interpolant F(x, y) was then evaluated regularly each 0.05$''$ with the routine E01SBF on a grid of $80\times80$ pixels to create the maps presented here. Only 37 fibers were considered to have enough signal for the definition of the photometric index (see Section~5.2), and these were also the ones used for the generation of the images. (They cover a hexagon centered on the fiber closest to the image peak). Figure~2 shows an example of such a reconstruction. The visual inspection of these images immediately allowed us to identify the loss of lock in the guiding system for four exposures, something confirmed later by the keywords in the headers of these images. 

The images were adjusted to 2D gaussians with the help of the task ``n2gaussfit" of IRAF.  It is well known that the seeing function does not fit a gaussian in the outer parts (wings). Therefore, we restrict these fits to a relatively small box of 0.8 arcsec (the estimated seeing). As a result of these fits we obtained the evolution of the following parameters: position (X, Y, position with respect to the central fiber),  fwhm (semi major axis), amplitude, ellipticity, and PA of the major axis. Note that for this particular application we are not interested in the absolute value of these variables (which depend on the assumptions about the function assumed), but only on their relative evolution along the night. 


\section{Results}


\subsection {Auxiliary Variables}

We studied the variation along the night of a set of auxiliary variables, which  will be compared with the photometric index defined in the next section. These variables were obtained from the ING meteorological data base, from the reduced spectra, and from the reduced images. As commented above, an important advantage of the IFS (especially the optical fiber approach) is the possibility to track independently the variations produced at the focal plane of the spectrograph (due, for instance, to instabilities of the mechanical structures, changes in the spectrograph focus, etc.) and those produced at the telescope focal plane (seeing changes, guiding instabilities, sky level variations, etc.).  Figures~3--5 show the changes of these variables along the night.

In Figure~3 we show the change in air temperature, humidity, atmospheric pressure, sky level, and airmass. In this figure we can see that the object was well located on the sky for these observations, and we could observe it for more than~7 hours at an airmass lower than~2.  The sky background level was somewhat unstable at the beginning of the night. In four images there was a relatively sudden drop ($\sim$  +100 minutes, in the time scale of the figure), but it was later discovered that for these images we lost the guiding system, so it was a spurious result (deviant results from these images were also found for other variables---see below).  
The background has a general trend to decrease over the night.  The last two points in the panel indicate  the increase due to the twilight. The values for the air~T, humidity, atmospheric pressure and wind intensity (not shown) indicate that the night had good stable conditions. Measurements for the seeing outside the dome were below 0.8 arcsec most of the night.    

Figure~4 shows a set of variables inferred from the spectra. The two lower panels illustrate the drifts of the spectra during the night along and across the dispersion directions (due to the motion of the image of the slit with respect to the detector). We can see that the spectra move at a relatively constant rate of 0.08~pix/h (across)  and 0.03~pix/h (along). Although these are very small changes, drifts of the spectra over the night make the observations sensitive to flat-field corrections, and extraction aperture location.  We did not find appreciable relative motion among the spectra, nor global rotation. Figure~4 also shows the evolution of the FWHM of the spectra integrated over the central band of the index (i.e.\ $f_2$, see below). This also shows a monotonic small change during the first part of the night, to reach a stable value for the second part of the night. The two deviant points (t $\sim$ +100~m.) correspond to the images for which the guiding system was lost (the relative change in the spectra intensity produces a small but appreciable change in the FWHM).  The variable denoted $\Delta$FWHM represents the relative change in the FWHM between the line ($f_2$) and the continuum ($f_1, f_3$) bands used for the definition of the index as it will be explained below. Although very small, we can appreciate a small trend. The variation of the total flux integrated over the 37 fibers and over the total spectral range used for the definition of the photometric index (i.e.\ $f_1 + f_2 + f_3$) is also represented, together with the individual $f_1, f_2,$ and $f_3$ fluxes. As we will see, the variation of these fluxes is clearly correlated with the star position in the bundle, seeing, and airmass. Finally, the top panel represents the ratio of the two continua used for the definition of the index (i.e.\ $f_1/f_3$). 

Figure~5 shows how the variables obtained after fitting the reconstructed images with a simple 2D Gaussian model change  over the night.  This figure shows a slight drift of the star image over the fiber bundle ($\sim$0.4~arcsec) during the first three hours until the point in which the auto-guiding was lost, to remain relatively constant the rest of the night. Also these panels show a clear correlation between the ellipticity and the FWHM, and an anticorrelation of these two variables with the amplitude of the gaussian fit and total flux. This behavior is expected due to the fact that, under these good seeing conditions, the fiber bundle undersamples the PSF.

\subsection {Index definition}

We define a photometric index probing the NaD lines as: 
\begin{equation}
I = \frac{f_1 - f_2 + f_3}{f_1 + f_2 + f_3},
\end{equation}
where $f_1, f_2$, and $f_3$ represent the sky subtracted flux (on a scale of photo-electrons)  in the following spectral ranges: 582.9200--588.5643, 588.5643--590.0887, and 590.0887--595.7321~nm, respectively.  Therefore,   the NaD feature is well centered within $f_2$, while $f_1$ and $f_3$ represent symmetrical blue and red continua, respectively.   Note that this index is insensitive to global (i.e.\ grey) variations in the flux.  Also note that by increasing the width of bands $f_1$ and $f_3$ (and assuming stable continuum values) one would reduce the standard deviation of a distribution of index values. This apparent gain in accuracy is obviously compensated by the fact that the index is less sensitive to changes in $f_2$. So, once the width of $f_1$ and $f_3$ are large enough compared with $f_2$ (so the statistical noise fixing the continuum is small compare to that due to $f_2$), no real gain is obtained by increasing the width of those bands (and further systematic errors can be introduced).  

As defined in eq.~(1), the index $I$ will probe the NaD lines under the assumption that $f_1$ and $f_3$ are stable continua.  Charbonneau et~al.\ (2002) have reported very stable continuum around the NaD lines, so one could only expect temporal variation of  $f_1$ and $f_3$ due to instrumental or atmospheric conditions. Some authors (e.g.\ Winn et~al., 2004 and references therein) have found relatively large time-dependent instrumental (blaze function) variations, which were interpreted as caused by flexure of the spectrograph. However, in the present case,  our fiber-fed spectrograph is mounted on a stable platform and, therefore, it is not affected by motions of the rotator. In any case, and in order to check instrumental stability, we have displayed in Figure~4 the fluxes in the bands $f_1$, $f_2$, and $f_3$ as a function of time. No apparent differences among these three plots can be directly observed, ruling out instrumental variations in flux of the order of 5--10 percent, as reported by other authors. In the top panel of Figure~4 we have also represented the ratio $f_1/f_3$, which has the expected much lower ($<$1~percent) variation with time (i.e.\ airmass) due the differential atmospheric extinction between the two bands.

From Eq.~(1) we can determined how relative changes in $f_2$ translate into relative changes in $I$. Specifically, 
\begin{equation}
\frac{\Delta I}{I} = \frac{-2(f_1 + f_3)f_2}{(f_1 + f_2 + f_3)(f_1 + f_3 - f_2)} \frac {\Delta f_2}{f_2}
\end{equation}
and, taking into account that  $f_1 \eqsim f_3 \eqsim  4f_2$, 
\begin{equation}
\frac{\Delta I} {I}   \eqsim - 0.25 \frac {\Delta f_2}{f_2}
\end{equation}
  
Thus, a relative change of the flux in the NaD band translates into a relative change $\sim$ 4 times smaller for the index, with the opposite sign.  So, taking into account that in a typical exposure we collect about  $7.15^. 10^6 e^-$ in $f_2$, the standard deviation due to photon noise for $f_2$ would be  $3.74^. 10^{-4}$ and, therefore, for the index  $I$ should be $\sim$  $9.5^. 10^{-5}$.

In addition to the index as defined above, we also evaluated a  {\it control\/}  index using the same definition as in Eq.~(1), but  a relatively line-free band for $f_2$.  Specifically the bands selected for $f_1$, $f_2$, and $f_3$ were: 590.300--595.146, 595.146--596.454, 596.454--601.300~nm, respectively.   

\subsection {Autocalibration}

The index as defined above was calculated for the 37 brightest spectra of each exposure. We excluded the four exposures for which we lost the the auto-guiding lock, and the four exposures taken at airmass larger than~2.  The resulting 5254 (37 spectra $\times$ 142 exposures) values are represented in Figure~6 (top) versus the total flux in the band (i.e.\ $f_t = f_1 + f_2 + f_3$).  Clearly there are some systematic deviations from the behavior expected in the case that noise were only due to photon noise.  This is particularly obvious  for the largest $f_t$ values, which are clearly non-linear.  At the lowest flux levels the deviations cannot be explained by the expected systematic trend in a ratio due to statistical errors. They are likely due to a small constant background due to scattered light  and, perhaps,  some non-linearity residuals too. In any case, these effects can be well calibrated using the data themselves (i.e.\ {\it autocalibration\/}). To this aim, we fit all the data points to a smooth function (splines3 of order~16) of the total flux, $I_{fit}(f_t)$, and we redefine the index as $I_c = I - I_{fit}(f_t)$. The corrected photometric index, $I_c$, for  the 5254 spectra are represented as a function of the total flux in Figure~6 (bottom).  These new $I_c$ values should be free of the non-linearity residuals commented above, since the dependence with $f_t$ has been removed .  

The distribution of values in Figure~6 was also used to check and select the proper methodology during the reduction process. For instance, we studied the effects of resampling the spectra to a common pixel-wavelength solution by analyzing the scatter in this figure. It was found that the scatter was appreciably larger when the spectra were resampled. This plot was also used to check the importance of the flat-field and linearity corrections. It was found that such corrections do indeed improve the quality of the results (i.e.\  the scatter in Figure~6 is reduced), but by relatively modest amounts.  

The next step was to obtain the photometric index value for each exposure (time),$I'$. This was done averaging the 37 contemporaneous individual values obtained in each exposure. We use an inverse variance weighting of the index (which is proportional to the total flux, $f_t$) in order to take into account  the statistical errors of each individual $I_c^i$ value. Therefore, 
\begin{equation}
I' = \frac{\Sigma I_c^i \times f_t ^i }{\Sigma f_t^i},
\end{equation}
where the subindex i runs over the 37 spectra obtained simultaneously in each exposure, and $f_t^i$ is the corresponding total flux (in photo-electrons). 

Note that apart from the possibility to perform the linearity calibration mentioned above, this procedure for generating the index is more advantageous than adding all the 37 spectra and generating the index from the combined spectrum, in terms of the influence of backgrounds (i.e.\ sky, scattered light, etc.) since these are actually weighted by the importance of each spectrum for the generation of the index at a given time. Also this method is also less sensitive to the actual spatial extend used (i.e number of fibers involved). Thus if, for instance, a number larger than the 37~fibers considered were used, the extra spectra would have had very little weight in $I'$, as a consequence of their low flux.   

Figures~7 and~8 show the dependence of the index with some of the auxiliary variables. In Table~1 the degree of correlation is indicated in 4th column.   Figure~7 and Table~1 show that the strongest correlation (0.76) is found against the airmass (which contrasts with the weak anticorrelation [$-$0.19], shown by the `{\it control\/} index', when $f_2$ is defined outside the NaD lines, as indicated in Section~5.2).   Initially we considered the possibility that this strong correlation between the index and the airmass could be due to geographical and temporal variations of the emission spectrum of the NaD lines produced in the Earth's atmosphere.  However, the emission spectrum has a very small contribution compared with the signal. In fact, considering a typical sky brightness of V~= 21.8~mag arcsec$^{-1}$ for La~Palma and an EW for the telluric emission of NaD of $\sim$150~\AA\ (Benn and Ellison 1998; Pedani 2004) , and the parameters used in our experiment (e.g.\  width of  $f_2\sim15$~\AA, fiber collecting area $\sim$0.16~arcsec$^{-1}$, etc.), one should collect more that $10^5$ photons from HD209458 for each photon coming from the NaD emission lines. On the other hand, Charbonneau et~al.\ (2002) have shown that the expected drop in the NaD band during the transit is $\sim$2.3$\cdot10^{-4}$ . So the amount we expect to detect during the transit is much larger (at least by one order of magnitude) than that due to the sky emission (the difficulty, of course, is in recording that signal  on top of the photon noise generated by the star).  Therefore, the strong correlation between the index and the airmass is likely to be dominated by the absorption produced by the earth atmosphere.  The fact that the index has a relatively strong correlation with airmass (which obviously change with time), may induce correlations with other variables which show a temporal variation, for which there is not a causal connection.  

\subsection {Decorrelation } 

By {\it decorrelation\/} we understand the process of removing trends in the index correlated with the auxiliary variables. 
This was based on a multiple-regression (least-square) fit to the data with a function which is linear in the coefficients of independent variables, following the prescriptions given by Bevington (1969).  Column~4th of Table~1 shows the correlation parameter for the 15 selected auxiliary variables. In order to have an idea of the effects of the decorrelation procedure reducing noise, we can compare the global standard deviation of the 142 data values before ($1.85\cdot10^{-4}$) and after ($1.11\cdot10^{-4}$) this procedure was applied.  Note that the standard deviation of the decorrelated values is remarkably close to the one expected in a purely photon-noise dominated regime. In fact, considering a typical level of counts per exposure of $3.1\cdot10^6$ ADU (counts) for $f_2$, a gain of 
$2.3~e^-/ADU$, and eq.~(3), a standard deviation of the index of $\sim$$0.95\cdot10^{-4}$ is expected in a photon-noise dominated regime. (It is interesting to note that the fact that the decorrelated signal shows a standard deviation close to that due to the photon noise suggests that the modal noise is very small or it can be well tracked by the selected variables).

As we can see in Table~1, the airmass is the variable which shows the highest correlation with the index. If the data are only decorrelated linearly against the airmass (i.e.\ if only the linear trend observed in the bottom panel of Fig.~7 is removed) the standard deviation is $1.20\cdot10^{-4}$. If we decorrelate against all other variables, but we exclude the airmass the standard deviation is 
$1.17\cdot10^{-4}$.  In Figure~9 (left) we show the original data (upper panel), the fit to a function linear to the airmass (center), and the decorrelated data (lower), as a funtion of time. Figures~9 (center) and (right) are similar but fitting the data  to a function linear to 14 (all but the airmass), and to all  the 15 selected auxiliary variables, respectively. It is interesting to note how the rest of the variables also carry information about the effects of the airmass.  

If we select for the decorrelation the six auxiliary variables derived from the spectra (background, shift~x, shift~y, FWHM, $\Delta$FWHM, total flux) the standard deviation is $1.39\cdot10^{-4}$. If we select the nine variables from the images (background, peak, 
shift~x, shift~y, distance to central fiber (111), FWMH, ellipticity, P.A., total flux) the standard deviation drops to $1.24\cdot10^{-4}$. This illustrates one of the advantages of using IFS: having control of the image characteristics at the focal plane offers the possibility to improve the photometry by removing correlated noise. 

Although this decorrelation method can be applied safely if the functional dependence of the auxiliary variables are different to that of the signal to be detected, it may destroy part of it when that is not the case.  This is more likely when many different variables are involved in the decorrelation process. Also the fact that our observations include only one transit, and this represents a relatively large fraction ($\sim$30 percent) of the whole observed period makes this possibility more likely. Ideally, observations during long out-of-transit periods will characterize more accurately the dependence on the variables. Obviously having data for multiple transits will minimize the possible artificial attenuation of the signal during the transit. In any case, to study the degree of destruction of the signal during the transit due to the decorrelation process we proceed as follows. Similarly to Brown et~al.\ (2002) we injected an artificial signal to the data (during the period that the transit took place) before the decorrelation was applied, and we analyzed how much of this signal was recovered after decorrelation. In a first step the effects of each auxiliary variable were analyzed individually. That was done comparing the input and output signals after decorrelating against each of the 15 variables. Column 5th in Table~1 indicates the fraction of signal recovered, when the decorrelation is done using only this variable. In an attempt to estimate the error for these conservation factors we used different levels of injected artificial signal (i.e.\ 0.5, 1. and 2~times the expected change according to the STIS--HST result), but the conservation factors were similar for these three cases. Table~1 shows that in several cases more than 
80~percent of the signal is recovered, while in other cases this is less than 30~percent. This is the case of three variables inferred from the spectra (shift~x, shift~y, fwhm). This is likely due to their monotonic behavior along the night, something that can mimic a transit occurred during the first part of the night as it was in the present case. However, a variable with a different type of dependence such as, for instance, the airmass (which decreases and increases along the night, with the inflexion point offset with respect to the transit) better preserves the signal. During the 132 minutes considered  as in-transit observations (i.e.\ excluding the ingress and egress) the airmass changed from 1.242 to 1.019. However, after the pass through the meridian,  the change in airmass from 1.019 to 1.242 (and on) was done out transit.   Then we studied the combined effects of several variables. We found that when all the 15 variables are used to remove noise (i.e.\ decorrelation process) most of the signal was actually destroyed: only 17~percent of the signal artificially injected is recovered after decorrelation. 

The next step in the analysis was to select a subset of auxiliary variables, which lead to a significant reduction of noise while preserving a substantial fraction of the signal.  After several trials based on the individual destruction factors, we found that three variables (airmass, FWHM of the image, distance to 111) give a good compromise.  Specifically, the standard deviation of the 142 points was 
$1.19\cdot10^{-4}$, while 79~percent of the signal is preserved after decorrelation. Although the addition of FWHM and distance to 111 implies a relatively modest improvement in the standard deviation with respect to the decorrelation against the airmass only, these variables were included because they take into account the position and shape of the image on the fiber bundle, which show a relatively strong correlation with the total flux (see Fig.~5). In any case, we checked that this selection did not affect our results, and values after decorrelating against the airmass only will also be given hereafter.  

At this point we evaluated the effects of the autocalibration (see Section~5.3) on the final standard deviation. If the decorrelation procedure is applied to the data without applying the autocalibration step a standard deviation of  $1.25\cdot10^{-4}$ is obtained, which compares with the value of $1.11\cdot10^{-4}$ mentioned above. (Values for which the total flux $f_t$ was larger than 
$1.18\cdot10^7$, were excluded since they show saturation---see Figure~6). This demonstrates that, the autocorrelation step allows to improve the photometric accuracy as well as to recover some of the saturated values (improving the temporal coverage, and the error of the mean value).         

\subsection {Time Series}

Figure 10 (upper panel) shows the time series data values. Vertical lines indicate the contacts. Following Charbonneau et~al.\ (2002) we define in-transit observations as those that occurred between second and third contact (i.e.\ $ |t-T_c| < 66.1$~m.), and out-of-transit those that occurred before first or after fourth contact (i.e.\ $|t-T_c| > 92.1$~m.). There are 45 in-transit observations and 84 out-of-transit observations, ignoring those observations that occurred during ingress and egress. 

As mentioned above the standard deviation of all the 142 values represented in Figure~10 is $1.19\cdot10^{-4}$. However, the standard deviation in-transit ($1.46\cdot10^{-4}$) is significantly larger than out-of-transit ($1.03\cdot10^{-4}$). We think this is attributable to the more unstable conditions of the atmosphere/background at the beginning of the night. Although our meteorological data do not show strong evidence for that (see Figure~3), variations of the telluric vapor absorption (e.g.\ Lundstrom et~al.\ 1991) and/or of the mesospheric sodium column density (e.g.\ Ge et~al.\ 1998) could produce these instabilities. In any case it is remarkable the fact that out-of transit, and considering data that expand for a period of $\sim$ 4 hours, the standard deviation is very similar to that expected from Poison-noise.  

The lower panel in Figure~10 shows the mean in-transit and out-of-transit values of the index: $(0.12 \pm 2.17)\cdot10^{-5}$ and $(0.06 \pm 1.12)\cdot10^{-5}$, respectively. The quoted 1-$\sigma$ errors correspond to the error in the mean of the statistical distribution 
(i.e.\ standard deviation/$\sqrt {n}$).  (If the index is decorrelated against the airmass only, the corresponding values are $[-0.04 \pm 2.16]\cdot10^{-5}$ and $[0.16 \pm 1.13]\cdot10^{-5}$, respectively.) These results are compatible with no variation of the index during the transit (i.e.\ the relative variation is $[0.06 \pm 2.44]\cdot10^{-5}$) and, taking into account equation~(3), no variation of the relative the flux in the NaD lines (i.e.\ $f_2$) with respect to the continuum during the transit. We also show in this figure the expected change in the index due to a change in $f_2$ as the one detected by STIS--HST (Charbonneau et~al.\ 2002). Provided that these authors found a relative  change in $f_2$ of  $(-2.32 \pm 0.57)\cdot10^{-4})$ during the transit, according to equation~(3) one should have expected a relative change in the index of $5.80\cdot10^{-5}$ with a conservation factor of~1, and $4.58\cdot10^{-5}$ with the current conservation factor of 0.79 (see Section~5.4).  (This neglects the slightly different width of the 12~\AA\  band used by Charbonneau et~al.\ with respect to the 15~\AA\  band used by us. This is justified by comparing their results for 12 and 38~\AA\ bands---see Table~2: a change of 3~\AA\ in the band width would imply a change $\sim 0.28\cdot10^{-5}$ in the index, about a factor of~5 smaller than the error quoted by them).  Therefore, if a signal with an amplitude as the one reported by Charbonneau et al.  were real, and after repeating the current experiment many times, the (gaussian) distribution of results should peak at $4.58\cdot10^{-5}$ with a sigma of 
$2.44\cdot10^{-5}$. However,  we (and Charbonneau et~al.) have performed the experiment only once, and the obtained result is about 2~$\sigma$ away from the expected value (for a signal as the one detected by Charbonneau et~al.\ data). If we take into account the quoted errors, the present data and the HST/STIS results disagree by more than 1~$\sigma$, but they are compatible within 
1.5~$\sigma$. 

This  limited statistical significance makes us cautious to over interpret the present results, which are about a factor~2 more uncertain than the HST data. Since atmospheric variables, and especially the airmass, dominate the decorrelation procedure, a better knowledge of the typical time scale for variation of the differential absorption in the NaD band by the Earth's atmosphere would be useful to further analysis. However, note that we could have obtained significantly better accuracy if the atmospheric conditions during the transit would have been similar to out-of-transit. In that case, the current  $2.44\cdot10^{-5} ($ 1-$\sigma$) error would have dropped to $1.9\cdot10^{-5}$. In addition, obviously, multiple transit observations would lead also to a significant reduction of this error. The recent ground based observations by Narita et~al.\ (2005) could not confirm nor contradict the Charbonneau et~al.\ result. Although these Subaru observations have much higher spectral resolution, its lower S/N and temporal resolution lead to an error in the predicted relative flux in NaD lines (for a 12~\AA\ band) larger (by a factor of about 4~times)  than that of the current observations.    

 \section{Conclusions}

The main conclusions of the present work can be summarized as follows:

1- We have shown that Integral Field Spectroscopy is a powerful technique for obtaining accurate relative time series photometry and, therefore, of great potential for studying transits of extrasolar planets. The three main advantages with respect to previously used methods  (based on imaging or standard slit spectroscopy) are: i)~it allows an  increase in the duty cycle of the observations (and, therefore, the S/N) by distributing the light into the two dimensions of the detector; ii)~the data can be {\it autocalibrated\/} from detector non-linearities and from background effects; and iii)~since the photometric index and  the image of the star are extracted from the same data-cube, the noise correlated with PSF characteristics can be removed.  Furthermore, the proposed IFS method should not have major demerits with respect to standard slit spectroscopy.    Although the present study is based on ground based observations, the full potential of IFS for these type of observations will be reached from space (e.g.\ JWST, TPF). 

2- These observations have also allowed us to explore the accuracy limits that can be achieved from the ground for these type of observations. In particular, using a photometric index that probes the strength of the NaD  lines, a standard deviation of  
$1.85\cdot10^{-4}$ was found (during a $\sim$7~hour period), which is about a factor~2 larger than that expected due to photon noise. However, after removing correlated noise by fitting the data to a function linear in three selected auxiliary variables the standard deviation drops to $1.19\cdot10^{-4}$, only  $\sim$25~percent larger than that due to photon noise. Simulations indicate that  
79~percent of the relative flux change during the transit should be preserved after decorrelating against those three variables.  Our nominal 1-$\sigma$ error for the mean in-transit (with respect to the mean out-transit value) for the index, is equivalent to 
$9.7\cdot10^{-5}$ for the relative flux in the NaD band ($f_2$) with respect to the continuum. This is substantially better than previous ground observations and only a factor $\sim$2~times worse than the 4-transit observations with STIS--HST. 

3- The present mean values for the photometric index in transit and out-of-transit are compatible with no extra depression of the NaD during the transit. Although the present results are in apparent contradiction with the HST--STIS results, we stress the limited statistical significance of this disagreement: they disagree at 1-$\sigma$, but they are consistent within 1.5-$\sigma$. We acknowledge that our results have larger nominal errors, and are more susceptible to systematically underestimate the signal through decorrelation. 

4-  The present study indicates that a good level of accuracy can be obtained from the ground with this type of IFS observations.  The 5254 spectra analyzed here were collected during only one night/transit  and, therefore, are insufficient to reduce the nominal errors to the level of the four transit STIS--HST data. Multiple transit observations will also allow to analyze the typical time scale variations for relevant atmospheric variables and, therefore, better understand the effects of the decorrelation procedure on the photometric index. 

5- From an instrument design perspective, this application of IFS can be fostered if:  i)~the fiber bundle (image slicer, array of micro-lenses) have an independent focussing mechanism, ii)~direct illumination of the detector is possible to generate flat fields, and iii)~the sky is simultaneously recorded. Fiber systems seem to have some generic advantages with respect to another type of IFU (e.g.\ azimuthal scrambling of the light within the fibers,  the possibility to use static spectrographs, etc.). Furthermore, our data indicate that modal noise or the temporal variation of the fiber throughput with wavelength are not significant sources of noise.   

Further multi-transit observations, with long out-of-transit data (ideally from space) should lead to the development of the full potential of the proposed method.  For ground based observations, the use of the recently developed new integral field spectrographs for 8m.\  class telescopes (e.g.\ GMOS: Allington-Smith  et~al.\ 2002, SINFONI/SPIFFI: Eisenhauer et~al.\ 2003, VIMOS: Le~Fevre et~al.\  2003) also opens an opportunity to push this technique forward.



\acknowledgments

We are grateful to 
Chris Benn, 
Ivo Busko,
Katrina Exter, 
Phill Hodge, 
Danny Lennon,  
Jes\'us Ma\'iz-Apell\'aniz,  and
Simon Tulloch, who have provided us with useful information. We also thank the anonymous referee for useful comments.   

This paper is based on observations made with the WHT, operated on the island of La Palma by the ING in the Spanish Observatorio del Roque de los Muchachos of the Instituto de Astrof\'isica de Canarias. We thank all the staff at the observatory for their kind support. 
Support for this work was also provided by the Spanish Ministry of  Education and Science through grant AYA2002-01055.

\begin{table}[h]
\begin{center}
\caption{Decorrelation Parameters\label{tbl-1}}

\begin{tabular}{r l c r c c }
\tableline\tableline
$\#$&\multicolumn{1}{c}{Variable} & Coefficient & \multicolumn{1}{c}{Error} & Correlation & Conservation Factor \\

\tableline

  1&  Airmass &      1.074 & 5.41E-04 & 0.76 & 0.88\\
  2& Total flux &     8.69E-04 &  1.33E-10  &\llap{$-$}0.14 & 0.88\\
  3& Background &      7.74E-02 & $-$6.61E-07 &\llap{$-$}0.48 & 0.75 \\
  4&  Spectra Shift (along disp.) &     2.11\phn &    4.64E-04 &0.38 & 0.33\\
  5& Spectra Shift (across disp.) &     1.60\phn & 3.60E-04  &\llap{$-$}0.29 & 0.32\\
  6& Spectra FWHM &      3.16\phn & $-$1.87E-05 &\llap{$-$}0.11 & 0.38\\
  7& Image Total Flux  &      5.82E-03 & $-$1.01E-08 &\llap{$-$}0.16 & 0.88\\
  8& Flux-ratio &      1.68\phn & $-$5.12E-04 &\llap{$-$}0.32 & 0.56\\
  9& Gaussian Peak &     4.55E-04 &  $-$6.08E-11 &0.27 & 0.77\\
  10& Gaussian X &     0.330 &   $-$2.81E-06 &8.9E-02 & 0.40\\
  11& Gaussian Y &      0.480  & 3.20E-05  &\llap{$-$}6.3E-02 & 0.46\\
  12& Distance to 111 &     0.384 &   3.24E-05  &0.17 & 0.99\\
  13& Gaussian FWHM &     0.422 &  $-$5.55E-05 &\llap{$-$}0.27 & 0.84\\
  14& Gaussian Ellipticity &     1.59\phn &   6.51E-04  & 2.3E-03 & 0.85\\
  15& Gaussian P.A. &     0.320 & $-$3.57E-06 &\llap{$-$}0.19 & 0.94\\
 \tableline
  & All & & & & 0.17\\
  & 1 +12+13 &&&& 0.79\\
\tableline
\end{tabular}
\end{center}
\end{table}

\begin{table}
\begin{center}
\caption{NaD measurements of HD209458b\label{tbl-1}}
\
\begin{tabular}{l r c l}
\tableline\tableline
\multicolumn{1}{c}{Work} & \multicolumn{1}{c}{Value} & Band (\AA) & Experimental setup \\

\tableline

Charbonneau et al.\ (2002) & $(23.2\pm5.7)\times10^{-5}$&  \phn12 &HST/STIS\\
                                         &  $(13.1\pm3.8)\times10^{-5}$ &   \phn38&  \\
                                         &  $(3.1\pm3.6)\times10^{-5}$&  100 & \\
  Narita et al. (2005)         &  $(30\pm40)\times10^{-5}$ &   \phn12 & Subaru/HDS\\
  Present work                  &  $(0.2\pm9.8)\times10^{-5}$ & \phn15  & WHT/INTEGRAL\\
\tableline
\end{tabular}


\end{center}
\end{table}

\begin{figure}
\epsscale{.70}
\plotone{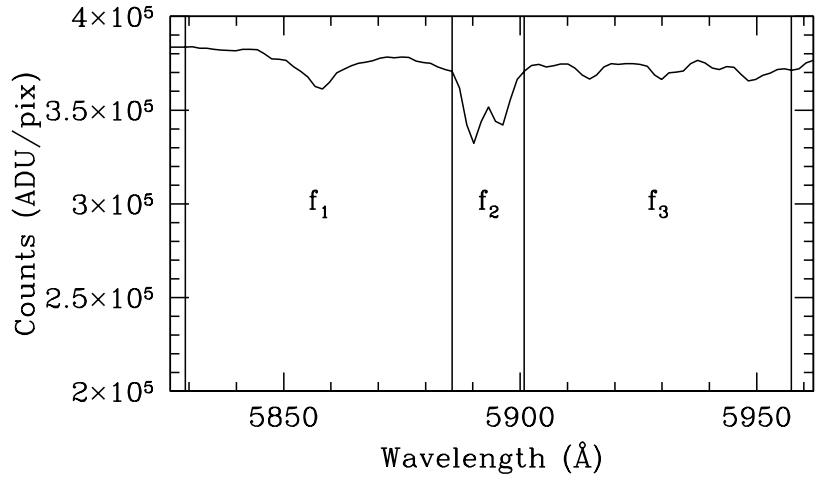}
\figcaption{Spectrum generated after combining all the reduced spectra obtained in a single 120 sec exposure. Vertical lines indicate the boundaries of the different bands used for the definition of the photometric index (see text).\label{fig1}}
\end{figure}

\begin{figure}
\epsscale{.50}
\plotone{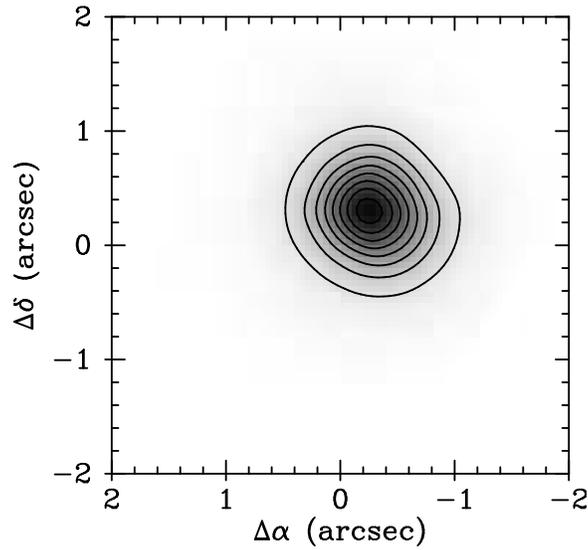}
\figcaption{Reconstructed image of the star at the focal plane during a typical 120 sec exposure (see text).\label{fig2}}
\end{figure}

\begin{figure}
\epsscale{.70}
\plotone{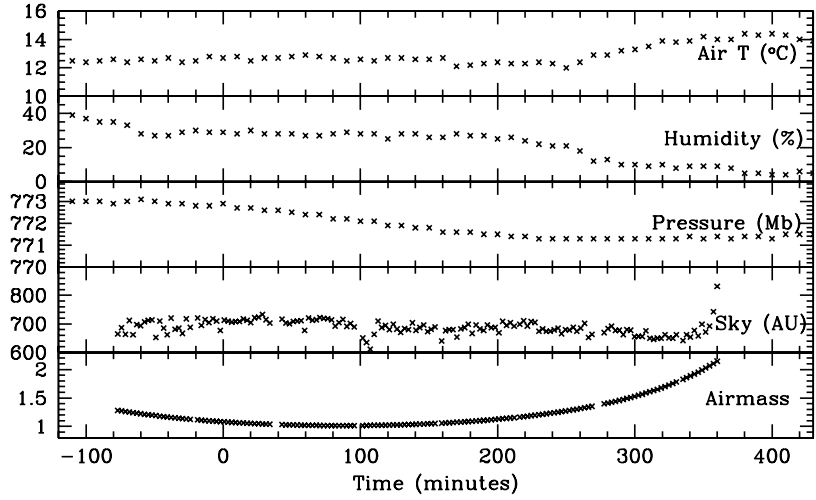}
\figcaption{Evolution of some atmospheric variables during the night (see text). Units for the horizontal axis are in minutes from the center of the transit.  AU refers to arbitrary units. \label{fig3}}
\end{figure}

\begin{figure}
\epsscale{.70}
\plotone{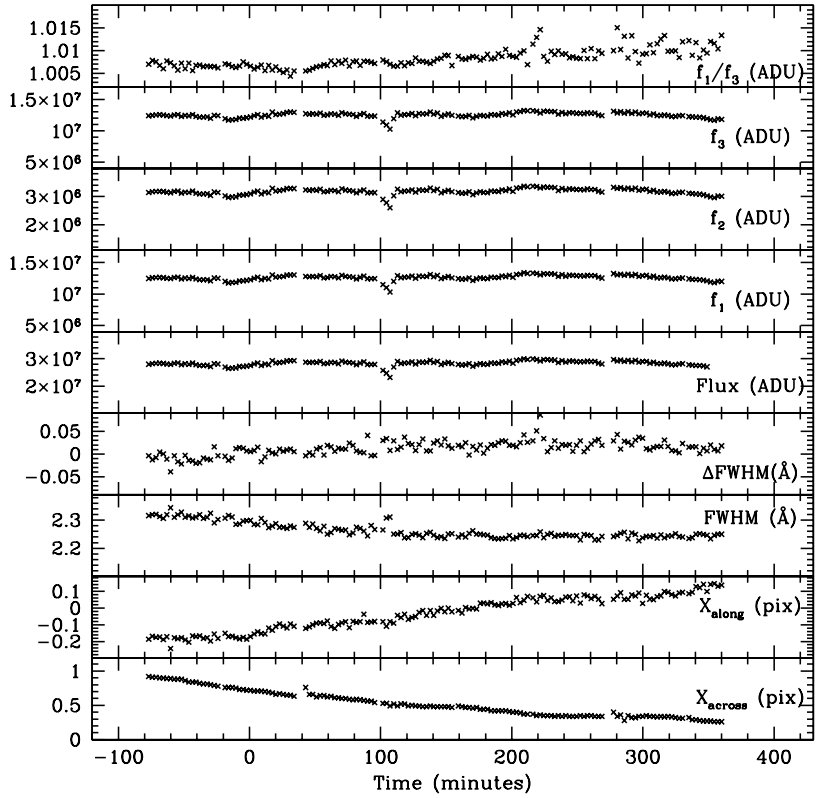}
\figcaption{Evolution of some variables inferred from the spectra during the night (see text). From bottom to upper panels: i)~Drift of the spectra in cross dispersion, ii)~shift of the spectra along the spectral direction, iii)~FWHM of the spectra, iv)~variation of the FWHM between the line and continua bands, v)~total flux contained in the 37 spectra considered for the definition of the photometric index and for the reconstruction of the images, vi), vii, and viii)~the corresponding individual fluxes for bands $f_1$, $f_2$, and $f_3$, and ix)~ratio $f_1/f_3$. Units for the horizontal axis are in minutes from the center of the transit. ADU refers to counts on the detector \label{fig4}}
\end{figure}

\begin{figure}
\epsscale{.70}
\plotone{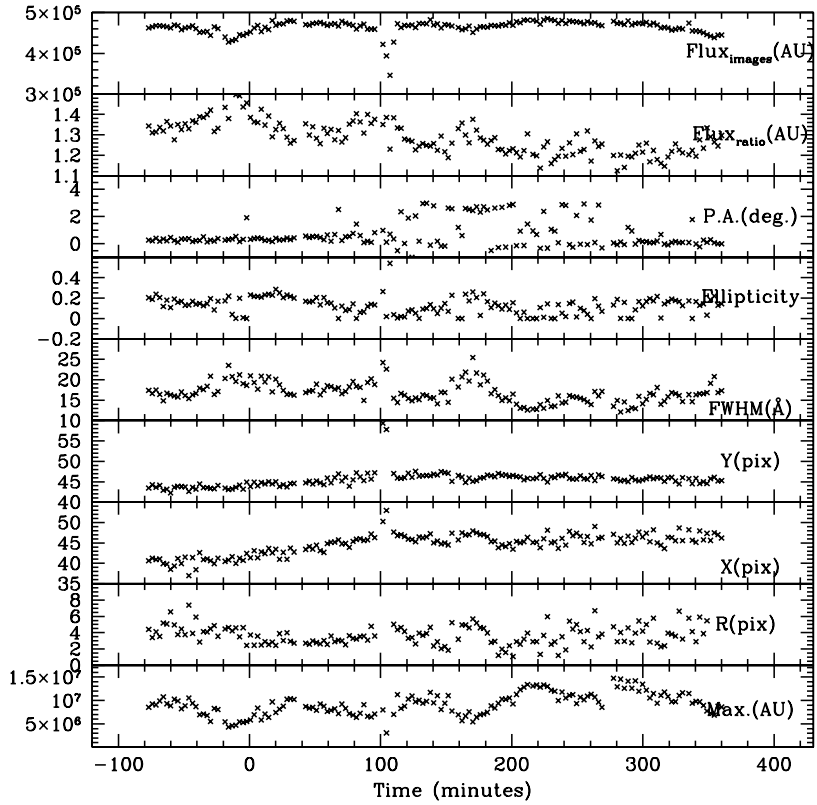}
\figcaption{Evolution during the night of some variables inferred from gaussian fits to the reconstructed images (see text). From bottom to upper panels: i)~value of the gaussian peak, ii)~distance to the fiber 111,  iii)~shift along the R.A.\ direction, iv)~shift along the DEC, v)~FWHM of the gaussian fit, vi)~ellipticity of the gaussian fit, vii)~P.A.\ of the major axis, viii)~ratio between the flux contained in the 
7~brightest fibers vs.\ that contained in the 37 selected fibers, and ix)~total flux in the images after interpolation. Units of vertical axis in ii), iii), and iv)~refer to pixels in the interpolated image (1~pixel~= 0.05~arcsec).  Units for the horizontal axis are in minutes since the center of the transit.  AU refers to arbitrary units.\label{fig5}}
\end{figure}

\begin{figure}
\epsscale{.70}
\plotone{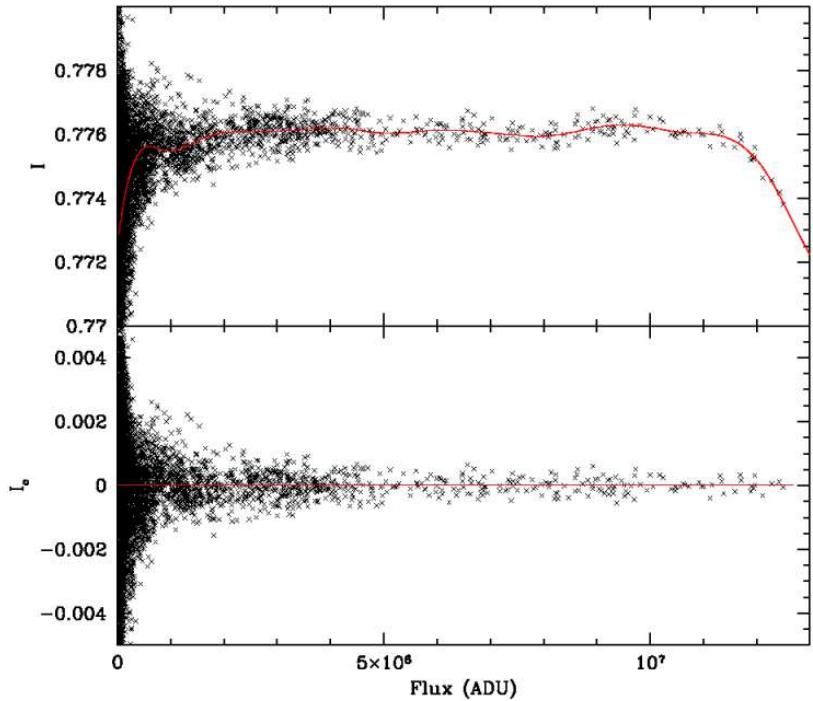}
\figcaption{Upper panel. Distribution of the photometric index as a function of the total flux (i.e.\ $f_t = f_1 +  f_2 + f_3$) for the 5254 spectra. Red line indicates the mean systematic behavior.
Bottom panel represent the same distribution after removing the mean systematic behavior. (see text)  \label{fig6}}
\end{figure}

\begin{figure}
\epsscale{.60}
\plotone{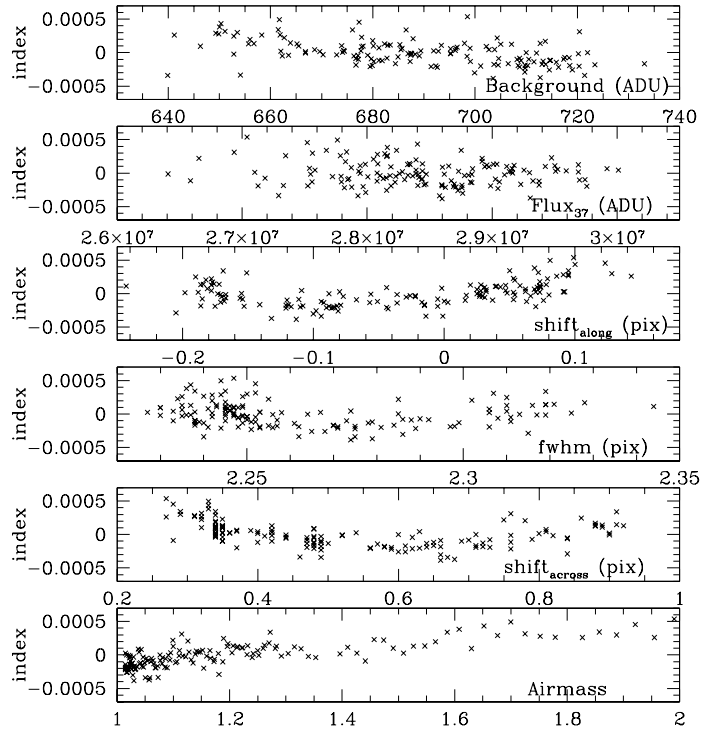}
\figcaption{Variation of the index $I_c$ as a function of several auxiliary variables. ADU refers to counts on the detector. \label{fig7}}
\end{figure}

\begin{figure}
\epsscale{.60}
\plotone{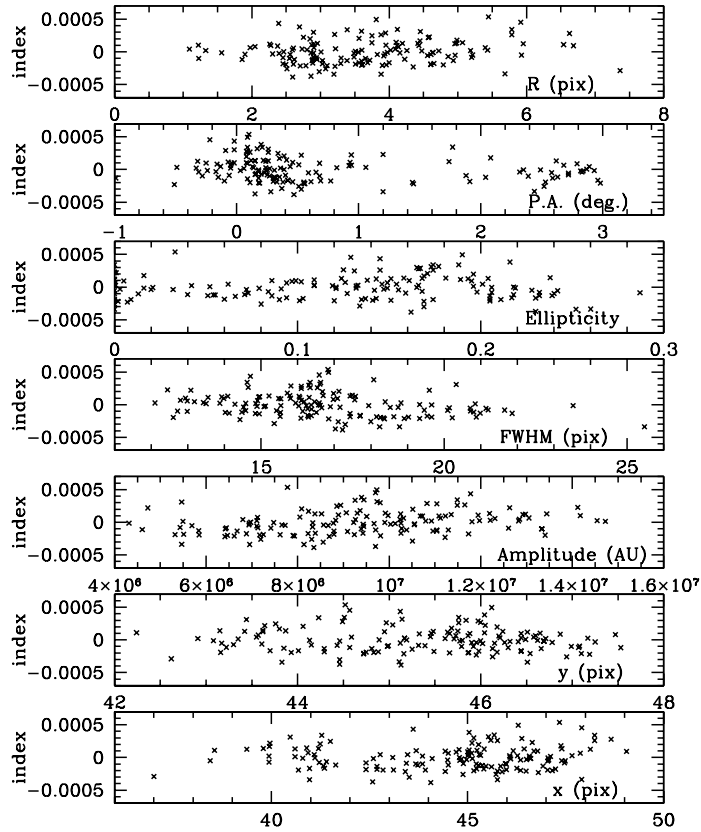}
\figcaption{Variation of the index  $I_c$ as a function of several auxiliary variables. The units of pixels indicated in several panels correspond to pixels in the reconstructed image, which have a scale of 0.05$''/pix$. R (upper panel) indicates the distance of the gaussian peak to the center of fiber number 111 (i.e.\ closest to the intensity peak). AU refers to arbitrary units. \label{fig8}}
\end{figure}

\begin{figure}
\epsscale{.70}
\plotone{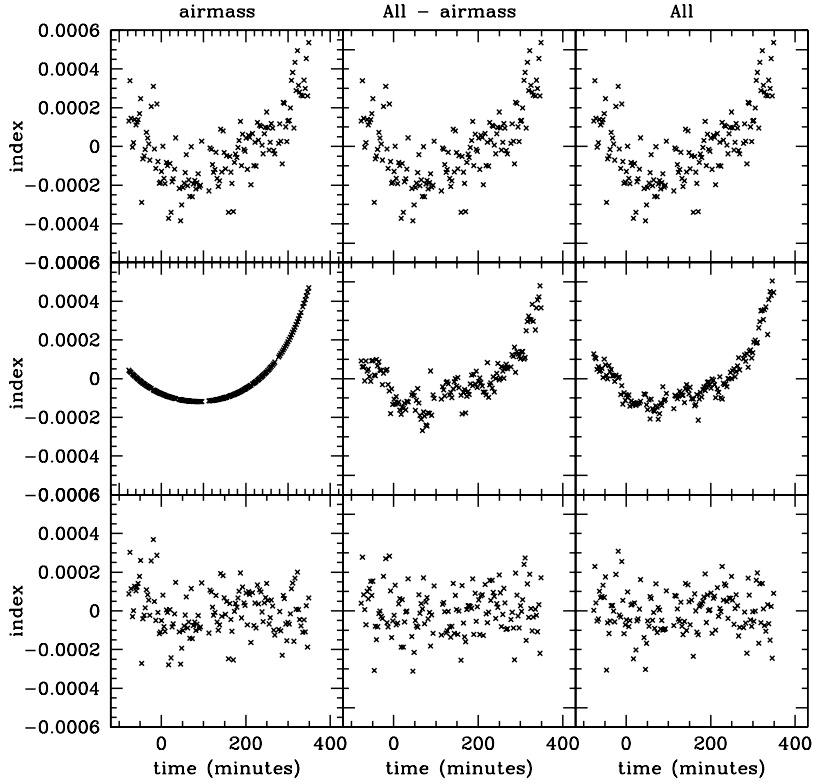}
\figcaption{Temporal variation of the index  $I_c$. The upper panels indicate the original distribution of values. Intermediate panels represent the fit of the original distribution to a function linear in the variables indicated at top (i.e.\ left: airmass; center: the 15 variables described in Table~1, excluding the airmass; and right: the 15 variables). The lower panels show the decorrelated distribution, which is obtained subtracting the fit to the original values. The standard deviation for these distributions are: $1.2\cdot10^{-4}$(left);  
$1.15\cdot10^{-4}$(middle): $1.1\cdot10^{-4}$(right).Units for the horizontal axis are in minutes from the center of the transit.  \label{fig9}}
\end{figure}

\begin{figure}
\epsscale{.60}
\plotone{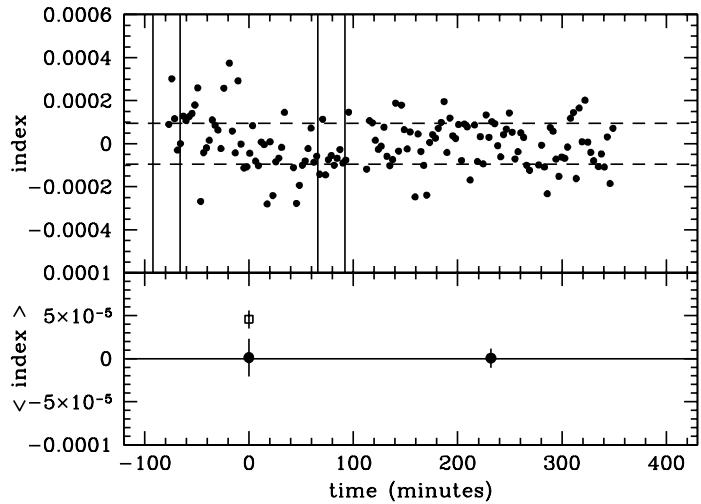}
\figcaption{Upper panel: Temporal variation of the index  $I_c$ after decorrelation with three selected variables (airmass, image FWHM, distance to 111). Vertical lines indicate (from left to right) first, second, third and fourth contact. Horizontal dashed lines show the expected $\pm$1-$\sigma$ according to photon noise. Lower panel shows the averaged values in-transit (i.e. between second and third contact) and out-of-transit (i.e.\ after fourth contact). The value corresponding to the HST observations by Charbonneau et~al.\  (2002) is shown with an open square (see text).  \label{fig10}}
\end{figure}






\clearpage

\end{document}